\documentclass[preprint,aps,showpacs]{revtex4}
\usepackage{epsfig}
\usepackage{float}
\RequirePackage{amssymb,amsmath}
\usepackage{makeidx}
\usepackage[latin1]{inputenc}
\usepackage{graphicx}
\usepackage{indentfirst}
\usepackage{color}
\begin{document}

\title{Induced photofission and spallation on the pre-actinide nucleus 
$^{181}$Ta.}
\author{A. Deppman}
\email{deppman@if.usp.br}
\affiliation{Instituto de Fisica, Universidade de S\~ao Paulo,
P. O. Box 66318, 05315-970 S\~ao Paulo, SP, Brazil}
\author{G. S. Karapetyan}
\email{ayvgay@if.usp.br}
\affiliation{Instituto de Fisica, Universidade de S\~ao Paulo, 
P. O. Box 66318, 05315-970 S\~ao Paulo, SP, Brazil}
\author{V. Guimar\~aes}
\email{valdirg@if.usp.br }
\affiliation{Instituto de Fisica, Universidade de S\~ao Paulo,
P. O. Box 66318, 05315-970 S\~ao Paulo, SP, Brazil}
\author{C. Gonzales}
\email{cgonzaleslorenzo@gmail.com}
\affiliation{Instituto de Fisica, Universidade de S\~ao Paulo,
P. O. Box 66318, 05315-970 S\~ao Paulo, SP, Brazil}
\author{A. R. Balabekyan}
\email{balabekyan@ysu.am}
\affiliation{Yerevan State University, Faculty of Physics, 
Alex Manoogian 1, Yerevan 0025, Armenia}
\author{N. A. Demekhina}
\email{demekhina@nrmail.jinr.ru}
\affiliation{Yerevan Physics Institute, Alikhanyan Brothers 2, 
Yerevan 0036, Armenia\\
Joint Institute for Nuclear Research (JINR), Flerov Laboratory of 
Nuclear Reactions (LNR), Joliot-Curie 6, Dubna 141980, Moscow region Russia}

\begin{abstract}
A study of photofission on $^{181}$Ta nucleus induced by bremsstrahlung 
photons with 
endpoint energies of 50 and 3500 MeV has been performed. The fission yields 
have been measured by using the induced-activity method in an off-line analysis.
The absolute photofission cross sections for the tantalum target at 50 and 
3500 MeV are found to be 5.4$\pm$1.1 $\mu$b and 0.77$\pm$0.11 mb, respectively,
and the corresponding deduced fissilities are (0.23$\pm$0.05)$\times 10^{-3}$ 
and (2.9$\pm$0.9)$\times 10^{-3}$. Mass- and charge-yield distributions were 
derived from the data. The results were compared with the simulated 
results from CRISP code for multi-modal fission by assuming symmetrical 
fission mode. 
\end{abstract}
\maketitle

\section{Introduction}

Induced fission has been studied over the years by a wide variety of 
projectiles and energy range. The information obtained from these experiments 
encompassed different characteristics of the fissioning system. 
Photofission is one of the most powerful tools for studying the fission 
process because of the well known spin 
selectivity of the excitation and the absence of the Coulomb barrier. 
Photons interact with nuclei by the quasideuteron mechanism and  
meson production. The main experimental problem in photofission studies is the 
lack of an intense source of monochromatic photon beam. Therefore, a large 
amount of photofission mass-yield distributions has been measured with 
bremsstrahlung spectra with a continuous energy distribution in the energy 
range of 300 to 1800 MeV \cite{Methasiri,Kroon,Lima,Emma}.  
Related to photofission on heavy nuclei, many works have been performed,
 with monochromatic photons at intermediate and high-energy regime, using
different experimental techniques  
\cite{Martins,Tavares1,Terranova1,Terranova2,Paiva}. In these experiments, 
the  photofission cross sections, and related fissility (the ratio of fission 
cross section to total nuclear photoabsoption cross section), were determined 
for pre-actinide targets with fissility parameter $Z^{2}/A<31.6$. At low photon 
incident energies ($\leq 30$ MeV), pre-actinide nuclei do not exhibit the 
resonance pattern characteristic of the giant resonance excitation  
\cite{Terranova1,Kroon}. Although very important for very heavy elements,
giant-resonance fission contribution is unimportant for elements
with atomic number $Z<83$ because of the high fission thresholds 
(22-27 MeV \cite{Tavares1}) and small values of the parameter $Z^{2}/A$.
For nuclei with mass number $A<210$ the trends of the photofission cross 
section show, conversely, an increase of several orders of magnitude, 
for incident energy from the fission threshold ($\sim$ 20-30 MeV) up to 
about 200 MeV \cite{Terranova1}. 

Mass-yield distribution for fission of different nuclear systems, which may 
exhibit symmetric and asymmetric modes and transition between them,
is of particular interest in the investigation of the fission process.
It is well known that mass distributions for actinide fission, induced 
by different projectiles and at intermediate energies, exhibit both symmetric 
and asymmetric modes. These fission modes can successfully be 
represented by  Gaussian curves, and less than five Gaussian curves
were adequate to describe mass distribution of 
induced fission on pre-actinides and higher 
actinides nuclei, where increasingly symmetric fission is observed 
\cite{Duijvestijn}. It is well known that fission is predominantly symmetric 
for pre-actinides with $A \leq 227$, since the targets are above the 
Businaro-Gallone point \cite{Businaro}. Actually, symmetric fission components 
have been observed in all photofission experiments with pre-actinides, but 
most of the time a quantitative comparison between each other is difficult
due to different methods used in the analysis.

The absence of a reliable theoretical model to predict fission yields, as 
well as the need for more experimental data,  motivated us to investigate 
photofission processes of pre-actinide nuclei. The essential goal of this 
paper is to present the measurement of the formation cross sections of fission 
fragments of $^{181}$Ta nucleus induced by bremsstrahlung photons with 
endpoint energies of 50 and 3500 MeV. In this experiment the total fission 
cross sections for both energies were derived from the experimental yields of 
fission fragments measured using off-line induced-activity method.
Investigation of photofission on such nucleus is also of importance for 
practical applications such as astrophysics, medicine, accelerator technology 
and nuclear waste transmutation.

Present experimental results for the fission process of the 
 $^{181}$Ta nucleus were compared with 
calculations by CRISP code based on the multi-modal fission approach 
\cite{Depp91}. This comparison allowed us to extract information on the 
reaction mechanism related to fission and spallation processes.

\section{Experimental Procedure}

The data for the photofission cross sections of $^{181}$Ta were obtained 
using bremsstrahlung photon radiation. 
The bremsstrahlung photon with endpoint energies of 3500 and 
50 MeV were obtained by using electrons of the Yerevan electron synchrotron 
and a linear accelerator of the injector type, respectively. The electrons 
were converted into bremsstrahlung-photon beam by means of tungsten 
converter of about 300 $\mu$m (about 0.1 radiation-length units) in thickness. 
The reaction chamber at the injector was arranged immediately after the 
converter.  A beam-cleaning and beam-formation system consisted of a set of 
collimators and it was used in the irradiation with 
high-energy photon beam. The high-energy photon beam passed through 
the first collimator, 3 $\times$ 3 mm$^{2}$ in dimension, and a cleaning 
magnet, which removed the charged component. The second collimator, 
10 $\times$ 10 mm$^{2}$ in dimension, was responsible for removing
the beam halo. 
The photon-beam intensity was measured by a Wilson-type quantometer
giving an average of $\sim 10^{11}$ equivalent 
quanta per second (eq. q. s$^{-1}$) at 3500 MeV and $\sim 10^{9}$ equivalent 
quanta per second (eq. q. s$^{-1}$) at 50 MeV. The photon beam 
intensities were evaluated from the monitor reactions 
$^{27}$Al($\gamma$, 2pn)$^{24}$Na and $^{65}$Cu($\gamma$, n)$^{64}$Cu, with 
known cross-sections \cite{Masumoto,Demekhina}. 
The 0.164g $^{181}$Ta target, with natural isotopic composition 
(99,98799$\%$) and  
0.0487 mm in thickness, was irradiated for 
196 min and 43 min with the  photon beam with endpoint energies 3500 MeV 
and 50 MeV, respectively. 

The yields of radioactive fragments were measured in an off-line 
analysis using a high-purity germanium (HpGe) detector (80 cm$^{3}$) 
with a resolution of about 0.2\% at the $^{60}$Co $\gamma$-transition energy of 
1332 keV. The $\gamma$-spectrometer detection efficiencies for four  
different target-detector distances, namely 0.0, 22.0, 7.0, and 25.0 cm, were 
determined by using the standard radiation sources of $^{22}$Na, $^{54}$Mn, 
$^{57,60}$Co, and $^{137}$Cs. To obtain the
energy dependence of the detector efficiency for energies above 1500 keV, 
we used also the data from the $^{27}$Al($\gamma$, 2pn)$^{24}$Na reaction 
(E$_{\gamma}$ = 2754 keV). The final energy dependence of the 
HpGe-detector efficiency was obtained with a precision of 10\%. 
Measurements of the $\gamma$-spectra started about 120 minutes after the 
completion of the irradiation and lasted a year.
The identification of the reaction products, and the determination of their 
production cross section, were performed considering the 
half-lives, energies and intensities of the $\gamma$-transition of 
the radioactive fragments.

The fragment production yields are considered direct and 
independent (I) in the absence of a parent isotope contribution 
(which may give a contribution via $\beta^{\pm}$-decays) and
are determined by the following equation:

\begin{eqnarray}
\hspace{-0.2cm}Y=\frac{\Delta{N}\;\lambda}{N_{p}\,N_{n}\,k\,\epsilon\,\eta\,(1-\exp{(-\lambda
t_{1})})\exp{(-\lambda t_{2})}(1-\exp{(-\lambda t_{3})})},\label{g1}
\end{eqnarray}

\noindent where $Y$ denotes the yields of the reaction fragment production; 
$\Delta{N}$ is the yield under the photopeak; $N_{p}$ is the projectile beam 
intensity (s$^{-1}$); $N_{n}$ is the number of target nuclei (in 1/cm$^{2}$ 
units); $t_{1}$ is the irradiation time; $t_{2}$ is the time of exposure 
between the end of the irradiation and the beginning of the measurement; 
$t_{3}$ is the time measurement; $\lambda$ is the decay constant (s$^{-1}$); 
$\eta$ is the intensity of $\gamma$-transitions; $k$ is the total coefficient 
of $\gamma$-ray absorption in target and detector materials, 
and $\epsilon$ is the $\gamma$-ray-detection efficiency.

If the yield of a given isotope receives a contribution from the 
$\beta^{\pm}$-decay of neighboring unstable isobars, the cross section 
calculation becomes more complicated \cite{Baba}. If the formation 
probability for the parent isotope is known from experimental data or 
if it can be estimated on the basis of other sources, then the independent 
cross sections of daughter nuclei can be calculated by the relation:

\begin{widetext}

\begin{eqnarray}
Y_{B}=&&\frac{\lambda_{B}}{(1-\exp{(-\lambda_{B}t_{1})})\,\exp{(-\lambda_{B}t_{2})}
(\,1-\exp{(-\lambda_{B} t_{3})})}\times\nonumber\\
&&\hspace*{-1.5cm}\left.\Biggl[\frac{\Delta{N}}{N_{\gamma}\,N_{n}\,k\,\epsilon\,\eta}-Y_{A}\,f_{AB}\,
\frac{\lambda_{A}\,\lambda_{B}}{\lambda_{B}-\lambda_{A}}
\Biggl(\frac{(1-\exp{(-\lambda_{A} t_{1})})\,\exp{(-\lambda_{A} t_{2})}\,(1-\exp{(-\lambda_{A} t_{3})})}
{\lambda^{2}_{A}}\right.\nonumber\\
&&\left.\qquad\quad\qquad\qquad\qquad-\frac{(1-\exp{(-\lambda_{B} t_{1})})\,\exp{(-\lambda_{B}
t_{2})}\,(1-\exp{(-\lambda_{B} t_{3})})}{\lambda^{2}_{B}}\Biggr)\right.\Biggr],
\end{eqnarray}
\end{widetext}
\noindent where the labels $A$ and $B$  refer to the parent and the daughter 
nucleus, respectively; the coefficient $f_{AB}$ specifies the fraction of $A$ 
nuclei decaying to a $B$ nucleus (this coefficient gives the information 
of how much the $\beta$-decay affects our data; and $f_{AB}=1$ is when the 
contribution from the $\beta$-decay corresponds to 100\%);
 and $\Delta{N}$ is the total photo peak yield associated with the 
decays of the daughter and parent isotopes. The effect of the forerunner 
can be negligible in some limit cases, for example, in the case where the 
half-life of the parent nucleus is very long, or in the case where the 
fraction of its contribution is very small. In the case when parent and 
daughter isotopes could not be separated experimentally, the calculated 
cross sections are classified as cumulative (C).

The yield of fission fragment production of Ta induced by photon beams with 
bremsstrahlung energy $E_{\gamma-max}=50$ and 3500 MeV are presented in Table I. 
In total, for the two energies measured, 61 yields were calculated for the 
fragment mass region $70 < A < 100$ u. The quoted uncertainties in the 
experimental yields are from the contributions of the statistical 
uncertainty ($\leq$ 2-3\%), uncertainty in the target thickness ($\leq$ 3\%), 
and uncertainty in the detector efficiency ($\leq$ 10\%).

\section{Results and Discussion}

From the production cross sections of the individual fragments we can 
construct the mass distribution (cross section of each isobar as a function 
of the mass number $A$). However, to obtain the cross section for each isobar 
with mass $A$ it is necessary to estimate the cross sections of the isotopes 
not measured by the induced-activity method. The cross sections for these 
fragments can be obtained from the analysis of the charge distribution of the 
corresponding isobar chain, i. e., the cross section as a function of $Z$ for 
a given $A$. We assumed that the charge distribution can be well described 
by a Gaussian function characterized by the most probable charge, $Z_P$,
(centroid of the Gaussian function) of an isobaric chain with mass $A$
and the associate width, $\Gamma_Z$. Moreover, the 
assumption is made that the most probable charge, as well as the width of the 
charge distribution, vary linearly with the mass of the fission fragment. 
The following parametrization of the production yield as a function of the 
charge of the fission fragment is adopted \cite{Duijvestijn}:

\begin{eqnarray}
Y_{A,Z}=\frac{Y_A}{\Gamma_Z\pi^{1/2}}\exp\left({-\frac{(Z-Z_P)^2}{\Gamma_Z^2}}\right),
\label{charge}
\end{eqnarray}

\noindent where $Y_{A,Z}$ is the independent yield of the nuclide ($Z,A$). 
The values $Y_A$ stands for the total isobaric yield for given mass number 
$A$, $Z_P$ is the most probable charge for the charge distribution 
of an isobars with mass number $A$ and $\Gamma_Z$ is the corresponding 
width parameter.
The values $Z_P$ and $\Gamma_Z$ can be represented as slowly varying linear 
functions of the mass numbers of fission fragments:

\begin{eqnarray}
Z_P&=&\mu_1+\mu_2A,\\
\label{most}
\Gamma_Z&=&\gamma_1+\gamma_2A,
\label{width}
\end{eqnarray}

\noindent where $\mu_1$, $\mu_2$, $\gamma_1$ and $\gamma_2$ are adjustable 
parameters determined by considering a systematic analysis of the fission
fragments. The obtained values for these parameters are listed
in Table II.

The mass distribution (isobaric yields) of fission fragments were then 
constructed by using the obtained values of $Y_A$ for each isobar chain. 
These mass distribution for the two endpoint energies are plotted in 
Figs. 1 and 2, respectively. In these figures the experimental data points 
of the present work is represented by black square symbol. In Fig. 2 we also
present the data from Ref.~\cite{Amroyan}, which includes 
data from spallation process of Ta target fragmentation 
($^{9}$Be, $^{22,24}$Na, and $A > 120$ u) at bremsstrahlung endpoint energy 
of 4 GeV (open circles).  

As one can see in both Figs. 1 and 2, data in the expected mass 
region for fission fragments agree with the assumption of a symmetric mass 
distribution. This mass region was, then, fitted with a Gaussian shape 
function given by: 

\begin{eqnarray}
Y_{f}=\lambda_{A}\exp\left({-\frac{(A-M_A)^{2}}{\Gamma_A^{2}}}\right)
\label{roldao}
\end{eqnarray}

\noindent where the parameter  $\lambda_{A}$ is the height, $M_A$ is the 
average mass number, and $\Gamma_A$ is the width of the Gaussian. 
These parameters were adjusted to the data and the obtained values are
listed in the Table II as $(M_A)_{exp}$ and $(\Gamma_A)_{exp}$ for each
endpoint energy.  The width, as well as the height, of the mass 
distribution clearly increase with increasing photon energy. From the 
mean value of the mass distributions, we can also conclude that, on average, 
three and six mass units are emitted before and after fission at low and 
intermediate energies, respectively.

The integration over the Gaussian gives the experimental fission yields 
for each endpoint energies. Actually, to get the fission yields, we  
had to multiply the Gaussian integration by a factor 0.5, to take into account
the double counting in the cross section due to the two fission fragments 
in each event. The experimentally determined values for fission yield for 
the endpoints energies 50 and 3500 MeV are $Y_{f}=$~5.4$\pm$1.1~$\mu$b 
and $Y_{f}=$~0.77$\pm$0.11~mb, respectively. The obtained measured total 
fission cross section at 50 MeV is in good agreement with the experimental 
value of 4.8$\pm$1.0 $\mu$b from Ref. \cite{Tavares1} for photofission 
of $^{nat}$Ta induced by 69 MeV monochromatic photons. 
For the higher energy photons (3500 MeV), the fission yield of the present 
work agrees well with the value of 0.64$\pm$0.06 mb obtained for reaction of 
bremsstrahlung with endpoint energy of 
3770 MeV on $^{181}$Ta \cite{Vartapetyan}.

From our data we could also estimate the fissility defined as the ratio of 
fission cross section to total nuclear photoabsoption cross section 
(D = Y$_{f}$/Y$_{abs}$). 
To determine Y$_{abs}$ it is necessary to take into account all possible 
decay channels of the excited nucleus being considered. 
The calculated fissility from our experimental fission cross section for 
the 50 MeV endpoint photons is $(0.23 \pm 0.05)\times 10^{-3}$ and for 
3500 MeV is $(2.9 \pm 0.5) \times 10^{-3}$. 
Here we considered the photoabsorption cross section for the endpoint 
energy 3500 MeV by taking the average values of data above the quasi-deuteron 
region of photonuclear absorption from Refs. \cite{Terranova2,Lima,Emma}. 
The fissility obtained for bremsstrahlung endpoint energy 
of 50 MeV is consistent with the trends calculated 
for photofission on $^{nat}$Ta induced by monochromatic photons of 69 MeV 
\cite{Tavares1} and for photofission of $^{181}$Ta at an incident 
monochromatic photon energy of 100 MeV \cite{Terranova1}. 
The fissility for the higher endpoint energy (3500 MeV) 
is in agreement with the systematics of fissilities as a function of 
$Z^{2}/A$ for photofission reactions with Ta 
targets at intermediate energies up to 6.0 GeV  \cite{Lima}. 
A general trend of increasing fissility with increasing photon energy for 
pre-actinide nuclei is consistent with what the results of
a systematic investigation of fission induced by bremsstrahlung photons on 
$Bi, Pb, Ti, Au, Pt, Os, Re, Ta$ and $Hf$ target nuclei  \cite{Ranyuk}.

\section{Calculation with CRISP code}

Calculations of fission cross sections within different models have provided 
good opportunities to estimate the validity of the various reaction mechanisms 
and to investigate characteristics of the processes taking place in reactions 
induced by different probes. Here we used the simulation code CRISP 
\cite{Depp2002a} to analyze our data.  CRISP is a Monte Carlo model code 
to describe nuclear reactions that uses a two step process 
\cite{Depp2002b,Depp2003}. First, an intranuclear cascade is simulated, 
following a time-ordered sequence of collisions in a many-body system 
\cite{Depp2004,Kodama,Goncalves1997}. When the intranuclear cascade 
finishes, the nucleus thermalizes and the competition between evaporation of 
nucleons and alpha-particles, and fission starts. This
code was recently used to analyse fission reactions induced by protons
 and photons \cite{Depp2013a,Depp2013b,Depp2013c}.
 
To analyze our data we pushed the code to simulate not only fission
process but also spallation reaction, which
might be an important reaction channel for proton and photon induced 
reaction at intermediate energies. The results of the CRISP calculations for 
both fission and spallation processes for our data on $^{181}$Ta target, 
at two endpoint energies, are shown in Figs. 1 and 2. 
The results of the simulation for fission 
calculation, given by the dotted line, show clearly that the
experimental distributions for both endpoint energies, taken into account 
only the intermediate mass (fission-fragment) region, can be well
reproduced by one symmetric Gaussian curve. Both peak position and width 
of the distributions are well described by the CRISP model. 
We used for fission calculation the experimental parameters of the charge 
distribution listed in Table II. 
For the mass distribution at endpoint energy of 50 MeV, the small fluctuations 
in the calculation is due to a limitation in the statistics. Although 
5 $\times$ 10$^{6}$ events were simulated, the resulting statistics for the 
fission of $^{181}$Ta is low due to the small fission yields. 
The total fission yields for $^{181}$Ta, as calculated by CRISP, are 
5.3 $\mu$b and 0.81 mb at low and intermediate energies, respectively, 
which agree completely with our experimental values obtained by a Gaussian 
fitting procedure described in the previous section. 
The calculated mean mass of the high
energy mass distribution, after evaporation of post-scission neutrons, is 
shifted to lower masses in comparison to the experimental ones. It means that 
the yields of fission fragments grow faster for higher energy, 
because of the considerable amount of high-energy photons.

The results of the CRISP calculation for the spallation process are indicated 
by dashed lines in the Figs. 1 and 2. The black solid line in Fig. 2 
indicates the sum of the calculated fission and spallation yields. 
As can be observed in Fig. 2, CRISP model does not give satisfactory results 
for the very light mass fragments region, and since nuclear fragmentation is 
not included in the model, this may be an indication that the fragmentation 
is relevant for explaining the fragment 
production in the mass range of 1-20 mass units.

CRISP calculations enables us to extract the fissility for 50 and 3500 MeV 
bremsstrahlung photon energies, which are $0.16 \times 10^{-3}$ and 
$0.41 \times 10^{-3}$, respectively. A qualitative agreement is obtained 
between the experimental and calculated fissility value for the low energy 
photofisison. For the high energy, the experimental fissility value is  
about one order of magnitude higher than the calculated value. 
A possible explanation for this is the fact that the 
total photon absorption yield is being overestimated by CRISP code due to a 
limitation of the model in taking into account all possible channels of decay 
of the excited target nucleus being considered.

Another source of information about the reaction dynamics that can 
be obtained from the simulation is the neutron production. The emission of 
neutrons starts already at the intranuclear cascade process with the 
pre-equilibrium production followed by the 
evaporation of neutrons from an equilibrated composite system. Both categories 
are referred to as pre-scission neutrons \cite{Strecker}. 
The post-scission neutrons are obtained when the system 
pass the scission point with neutron emission by the residual fragments.
The neutron production can then have the following contributions:

\noindent (i) from the composite system; 

\noindent (ii) during the transition of the composite system through the 
saddle-point configuration towards the scission point;

\noindent (iii) during the neck rupture; 

\noindent (iv) from the accelerating fragments and; 

\noindent (v) after completion of their acceleration.
\par\hfill

The contributions from (i), (ii) and (iii) are not distinguishable and are 
therefore considered as pre-scission neutrons. The contributions from 
(iv) and (v) are classified as post-scission neutrons.
With the CRISP model we can also obtain the average number of pre- and 
post-scission emitted neutrons. We present in Table III the average fissioning 
nucleus mass $A_f$, the average mass of fission 
fragment mass distribution after evaporation, $A_{ff}$, as well as the 
average number of pre- and post-scission neutrons.  The sum of the two 
neutron emission contributions gives the total number of emitted neutrons, 
which can be compared with the experimental values 
in Table III. We observe a good agreement between the calculated and 
the experimental values for the low energy induced fission, showing that 
the theoretical predictions for the emission of neutrons are correct. 
However, for the higher energy (3500 MeV) the 
calculations of the neutron multiplicities from the excited $^{181}$Ta nucleus 
is somewhat overestimated. The experimental neutron emission is 11 neutrons 
while the calculation gives 16 neutrons.  Again we emphasize 
that the larger is the intranuclear cascade, and the evaporation/fission 
chains, more difficulty is the calculation. In this case, however, it is 
possible to observe that the main contribution to the disagreement between 
calculation and experiment comes from the fission fragment evaporation
(post-scission neutrons).  There are some points in the calculation which 
could lead to this discrepancies, but most of them are also present in the 
evaporation of the compound nucleus formed after the intranuclear cascade. 
There is one particular mechanism  which is related only to the fragment 
evaporation, namely, the distribution of the excitation energy between the 
two fission fragments. 
In the CRISP model it is assumed that the excitation energy of the 
fissioning nucleus will be distributed to the two fragments proportionally 
to their masses, keeping the total excitation energy constant. Behind this 
assumption is the idea that there is no energy transfer in the scission 
process from microscopic to collective degrees of freedom. This is not 
necessarily true, and the large number of neutrons evaporated from the 
fragments may be an indication that part of the excitation energy may 
appears as collective motion of the fragments.
We are working on this issue to improve the simulation code CRISP, but
we can say that the present analysis with the CRISP code already 
indicates that our theoretical model gives a good description of the 
dynamical process taking place inside the nucleus during reactions 
at intermediate energies.

\section{Conclusion}

In this work we present the results of the investigation of 
the induced fission of $^{181}$Ta nucleus by bremsstrahlung photon beams 
with endpoint energies of 50 and 3500 MeV.  Photofission yields have been 
measured taking advantage of the induced-activity method in an off-line 
analysis.
The absolute photofission yields have been determined for the two very 
different energy regimes taking into account the photon spectrum
measured. Photofissility values were subsequently deduced 
for each endpoint energy of photons. The obtained total fission yields and 
fissility values have been found to agree quite well with 
the values obtained from the measurement at 69 and 3770 MeV of incident 
photon energies. 
An analysis of the charge and mass distribution of fission fragments 
from $^{181}$Ta target have been performed with the CRISP code. The 
comparison between calculated parameters for $^{181}$Ta target and 
the experimental data has shown that the CRISP model gives a good 
description  of the main characteristics of the reaction 
under investigation at the two endpoint energies (50 and 3500 MeV). 
The small disagreement 
between experimental and calculated values was found for the 
neutron evaporation from the hot fission fragments. We argue that this 
problem can be related to the transfer of energy from microscopic to 
collective degrees of freedom in the fissioning system, which is being
improved in the CRISP code.

\section*{Acknowledgment} G. Karapetyan is grateful to Funda\c c\~ao de 
Amparo \`a Pesquisa do Estado de S\~ao Paulo (FAPESP) 2011/00314-0 and 
2014/00284-1, and also to International Centre for Theoretical Physics (ICTP) 
under the Associate Grant Scheme. A. D.  and V. G. acknowledges the support 
from CNPq under grant 305639/2010-2  and 302969/2013-6, respectively.

\medbreak

\newpage
\begin{table}
\small
\caption{Yields of fission fragments measured for the reaction with photons 
at E$_{\gamma-max} $=50 and 3500 MeV on $^{181}$Ta target.}
\tiny
\begin{tabular}{|c|c|c|c|}
\hline
Element&Type& \multicolumn{2}{|c|}{Yield, $\mu$b/eq.q. }\\
\hline
& &E$_{\gamma-max} $=50 MeV&E$_{\gamma-max} $=3500 MeV\\ 
\hline 
$^{59}$Fe  & C &  -            & $\leq$10.0\\
$^{64}$Cu  & I &  -            & 12.0$\pm$2.0\\
$^{65}$Zn  & C &  -            & 18.0$\pm$5.0\\
$^{69m}$Zn & I &  -            & 21.0$\pm$3.0\\
$^{71m}$Zn & C &  -            & 22.0$\pm$2.0\\
$^{72}$Zn  & C & 0.13$\pm$0.02 & 23.0$\pm$3.0\\
$^{72}$Ga  & I &  -            & $\leq$7.2\\
$^{73}$Ga  & C & 0.14$\pm$0.02 & 25.0$\pm$3.7\\
$^{74}$As  & I & 0.20$\pm$0.03 & 20.0$\pm$4.0\\
$^{75}$Se  & C & -             & 29.0$\pm$4.0\\
$^{76}$As  & I & 0.20$\pm$0.03 & 22.0$\pm$3.0\\
$^{77}$Ge  & C & 0.20$\pm$0.03 & 29.0$\pm$4.0\\
$^{77}$Br  & I &  -            & $\leq$5.0\\
$^{78}$Ge  & C & 0.25$\pm$0.04 & 22.0$\pm$2.2\\
$^{78}$As  & I &  -            & 15.0$\pm$2.3\\
$^{82}$Br  & I &  -            & 23.0$\pm$4.0\\
$^{84}$Br  & C & 0.40$\pm$0.06 & 31.0$\pm$6.0\\
$^{84}$Rb  & I &  -            & 4.0$\pm$0.7\\
$^{84m}$Rb & I &  -            & $\leq$5.0\\
$^{85m}$Sr & C &  -            & 27.0$\pm$5.0\\
$^{86}$Rb  & I & 0.36$\pm$0.05 & 15.0$\pm$3.0\\
$^{87}$Kr  & C & 0.45$\pm$0.07 & 33.0$\pm$3.3\\
$^{87}$Y   & C &  -            & $\leq$8.8\\
$^{88}$Kr  & C & 0.40$\pm$0.06 & 30.0$\pm$6.0\\
$^{88}$Y   & I & -             & $\leq$7.0\\
$^{88}$Zr  & C & -             & $\leq$10.0\\
$^{90m}$Y  & C & -             & 41.0$\pm$6.0\\
$^{91}$Sr  & C & 0.38$\pm$0.06 & 35.0$\pm$7.0\\
$^{91m}$Y  & C & -             & $\leq$9.0\\
$^{92}$Sr  & C & 0.36$\pm$0.05 & 33.0$\pm$7.0\\
$^{92}$Y   & I & -             & $\leq$7.0\\
$^{93}$Y   & C & 0.29$\pm$0.04 & 36.0$\pm$6.0\\
$^{95}$Zr  & C & 0.40$\pm$0.06 & 34.0$\pm$5.0\\
$^{95m}$Nb & I & -             & 9.0$\pm$1.4\\
$^{96}$Nb  & I & -             & 32.0$\pm$5.0\\
$^{96}$Tc  & I & -             & $\leq$4.0\\
$^{97}$Zr  & C & 0.28$\pm$0.04 & 27.0$\pm$5.4\\
$^{99}$Mo  & C & 0.27$\pm$0.04 & 29.0$\pm$6.0\\
$^{100}$Pd & C & -             & 13.0$\pm$2.6\\
$^{102}$Rh & C & -             & 22.0$\pm$5.0\\
$^{103}$Ru & C & 0.15$\pm$0.03 & 25.0$\pm$5.0\\
$^{105}$Ru & C & 0.15$\pm$0.03 & 13.0$\pm$2.6\\
$^{105}$Rh & I & -             & $\leq$4.0\\
\hline
\end{tabular}
\end{table}

\newpage
\begin{table}
\caption{Parameters values obtained for the mass and charge distributions 
for $^{181}$Ta target at the endpoint energies of 50 and 3500 MeV.}
\begin{tabular}{|c|c|c|} \hline
Parameter	   & 	50 MeV             & 	3500 MeV \\ \hline
($\lambda_A)_{exp}$ & 0.00040$\pm$0.00001   &   0.0410$\pm$0.0002\\
($\lambda_A)_{cal}$ & 0.0004	           &	   \\
($M_A)_{exp}$	   & 88.0$\pm$0.6	   &   85.0$\pm$0.6\\
($M_A)_{cal}$	   & 87.9	           &   82.5	\\
($\Gamma_A)_{exp}$  & 14.39$\pm$0.20	   &   23.5$\pm$0.3\\
($\Gamma_A)_{cal}$  & 15.0		   &	   \\
($\mu_1)_{exp}$	   & 1.69 $\pm$0.11	   & 0.779$\pm$0.046\\
($\mu_1)_{cal}$	   & 1.690	           &	  \\
($\mu_2)_{exp}$	   & 0.397$\pm$0.002	   & 0.420$\pm$0.001\\
($\mu_2)_{cal}$	   & 0.397		   &	  \\
($\gamma_1)_{exp}$  & 0.590$\pm$0.007	   & 0.590$\pm$0.003\\
($\gamma_1)_{cal}$  & 0.59		   &	   \\
($\gamma_2)_{exp}$  & 0.0050$\pm$0.0009	   & 0.0050$\pm$0.0004\\
($\gamma_2)_{cal}$  & 0.0050		   &	   \\
\hline
\end{tabular}
\end{table}

\newpage
\begin{table}
\caption{Calculated and experimental parameters obtained for the mass 
distribution: mean mass of the fissioning nucleus [$(A_f)_{cal}$ ]
after evaporation of pre-scission neutrons from the compound nucleus; 
mean mass of the fissioning nucleus [$(A_{ff})_{cal}$] after evaporation of 
post-scission neutrons from fragments; experimental mean mass of the 
fissioning nucleus [$(A_{ff})_{exp}$], which includes both type of evaporated 
neutrons; number of pre- and post-scission neutrons, evaporated from the 
excited nucleus, fission cross sections.}
\begin{tabular}{|c|c|c|}  \hline
Parameter                       & 50 MeV              & 3500 MeV    \\ 
\hline 
$(A_{ff})_{exp}$                  & 176.0 $\pm$0.9      & 170.0$\pm$0.9    \\
$(A_{ff})_{cal}$                  & 175.8               & 165.0    \\
$(A_f)_{cal}$                    & 180.4               & 178.3     \\
(pre-scission neutrons)$_{cal}$  & 0.6                 & 2.7       \\
(post-scission neutrons)$_{cal}$ & 4.6                 & 13.3       \\
(evaporated neutrons)$_{exp}$    & 5.0$\pm$ 4.0        & 11.0$\pm$1.2   \\ 
(fission cross section)$_{exp}$ & 5.4$\pm$ 1.1 $\mu$b & 0.77$\pm$0.11 mb \\
(fission cross section)$_{cal}$ & 5.3 $\mu$b          & 0.81 mb \\
\hline
\end{tabular}
\end{table}

\newpage
\begin{figure}
\caption{Mass-yield distribution for photofission of $^{181}$Ta 
at endpoint energy of 50 MeV. The present experimental data are shown by 
solid square symbol, the results of CRISP code calculation for fission is given
by the dotted line curve, and for spallation by the dashed line curve.}
\epsfig{file=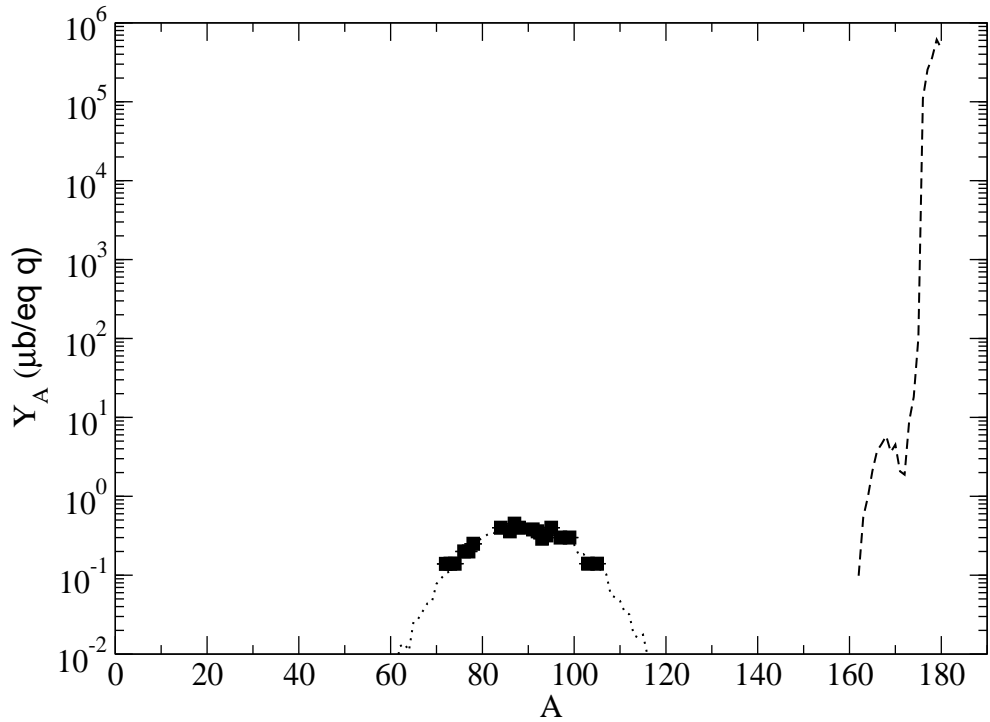,height=15cm,width=12cm,angle=-90.}
\end{figure}

\newpage
\begin{figure}
\caption{Mass-yield distributuion of photofission of $^{181}$Ta 
at endpoint energy of 3500 MeV. The present experimental data are shown by 
the solid square symbol. The open circle symbol correspond to data
taken from Ref.~\cite{Amroyan} wich includes also spallation contribution.
The results of the CRISP code calculation for 
fission is given by the dotted line, and for spallation by the dashed line
curve. The black solid line indicates the sum of the calculated fission and 
spallation yields.}
\epsfig{file=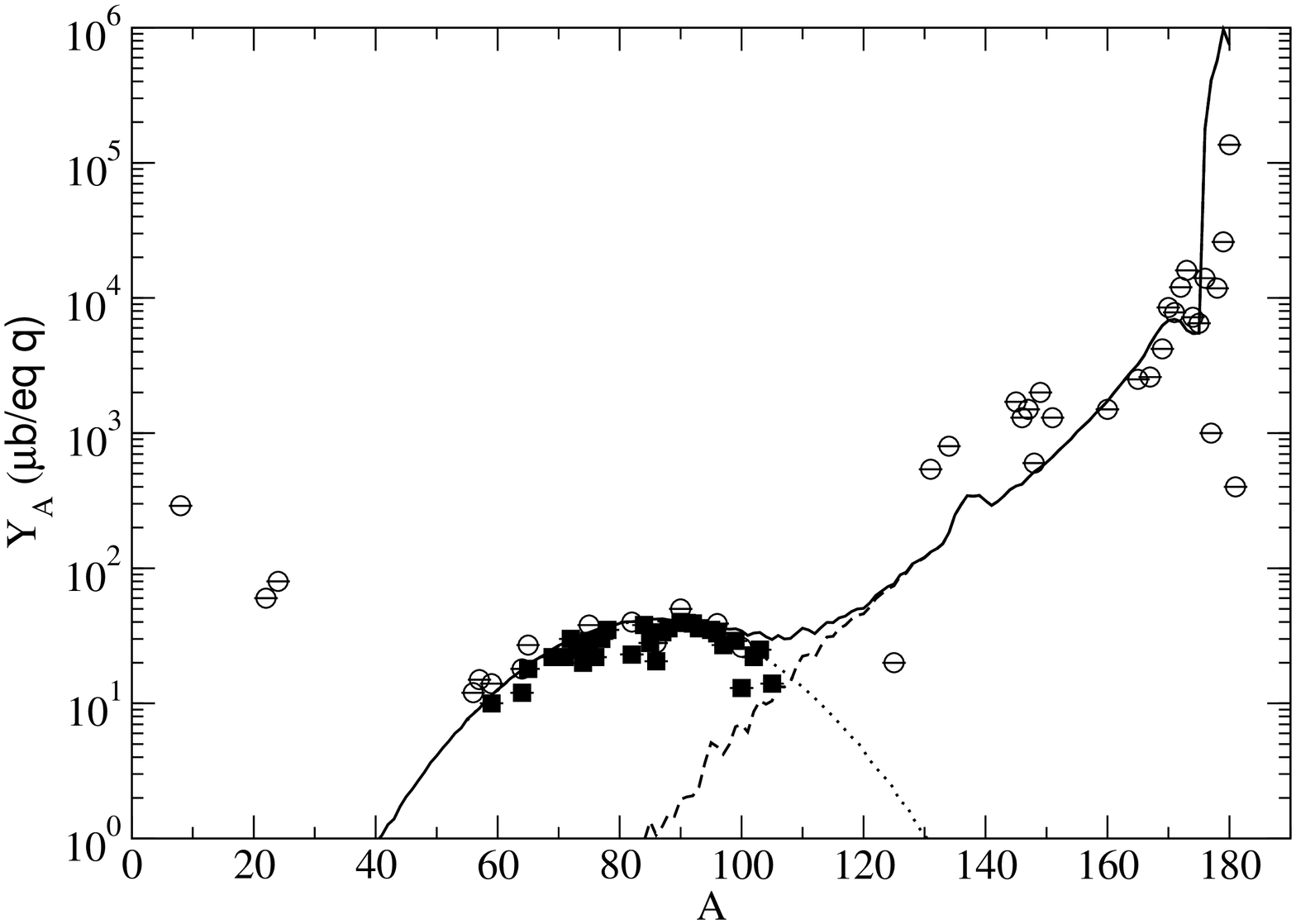,height=15cm,width=12cm,angle=-90.}
\end{figure}


\bigskip

\begin{thebibliography}{99}

\bibitem{Methasiri} T. Methasiri and S. A. E. Johansson, Nucl. Phys. A 
{\bf167}, 97 (1971).

\bibitem{Kroon} I. Kroon and B. Forkman, Nucl. Phys. A {\bf197}, 81 (1972).
\bibitem{Lima} D. A. De Lima, J. B. Martins and O. A. P. Tavares, 
IL Nuovo Cim. A {\bf103}, 701 (1990).
\bibitem{Emma} V. Emma, S. Lo Nigro and C. Milone, 
Nucl. Phys. A {\bf257}, 438 (1976).
\bibitem{Tavares1} O. A. P. Tavares, J. B. Martins, E. de Paiva {\it et al.}, 
J. Phys. G: Nucl. Part. Phys. $\bf19$, 805 (1993).
\bibitem{Martins} J. B. Martins, E. L. Moreira, O. A. P. Tavares {\it et al.}, 
IL Nuovo Cim. A $\bf101$, 789 (1989).
\bibitem{Terranova1} M. L. Terranova, O. A. P. Tavares, G. Ya. Kezerashvili 
{\it et al.}, J. Phys. G: Nucl. Part. Phys. {\bf22}, 511 (1996).
\bibitem{Terranova2} M. L. Terranova, G. Ya. Kezerashvili, A. M. Milov 
{\it et al.}, J. Phys. G: Nucl. Part. Phys. {\bf24}, 205 (1998).
\bibitem{Paiva} E. de Paiva, O. A. P. Tavares, M. L. Terranova, 
J. Phys. G: Nucl. Part. Phys. {\bf 27}, 1435 (2001).
\bibitem{Duijvestijn} M. C. Duijvestijn, A. J. Koning {\it et al.}, 
Phys. Rev. C $\bf59$, 776 (1999).
\bibitem{Businaro} U. L. Businaro and S. Gallone, Nuovo Cimento {\bf 1}, 
629 (1955).
\bibitem{Depp91} A. Deppman, S. B. Duarte, G. Silva {\it et al.}, 
J. Phys. G $\bf30$, 1991 (2004).
\bibitem{Masumoto} K. Masumoto, T. Kato, and N. Suzuki, 
Nucl. Instrum. Methods Phys. Res. $\bf157$, 567 (1978).
\bibitem{Demekhina} N. A. Demekhina and A. S. Danagulyan, Yad. Fiz. $\bf24$, 
681 (1976) [Sov. J. Nucl. Phys. $\bf24$, 355 (1976).
\bibitem{Baba} H. Baba, J. Sanada, H. Araki {\it et al.}, 
Nucl. Instrum. Methods A $\bf416$, 301 (1998).
\bibitem{Amroyan} K. A. Amroyan, S. A. Barsegyan, N. A. Demekhina, Sov J. Nucl. Phys. {\bf 56}, 4 (1993).
\bibitem{Vartapetyan} G. A. Vartapetyan, N. A. Demekhina, V. I. Kasilov, 
{\it et al.}, Sov J. Nucl. Phys. {\bf 14}, 37 (1972).
\bibitem{Ranyuk} Yu. N. Ranyuk and P. V. Sorokin, J. Nucl. Phys. (USSR) {\bf 5}, 37 (1967) (Engl. Transl. Sov J. Nucl. Phys. {\bf 5}, 26 (1967).
\bibitem{Depp2002a} A. Deppman, O. A. P. Tavares, S. B. Duarte, {\it et al.}, 
Comp. Phys. Comm. ${\bf145}$, 385 (2002).
\bibitem{Depp2002b} A. Deppman, O. A. P. Tavares, S. B. Duarte, 
{\it et al.}, Phys. Rev. C ${\bf66}$, 067601 (2002).
\bibitem{Depp2003} A. Deppman, O.A.P. Tavares, S. B. Duarte, {\it et al.}, 
Nucl. Instr. Meth. B  ${\bf211}$, 15 (2003).
\bibitem{Depp2004} A. Deppman, S. B. Duarte, G. Silva {\it et al.}, 
J. Phys. G: Nucl. Part. Phys. ${\bf30}$ 1991 (2004).
\bibitem{Kodama} T. Kodama, S. B. Duarte, K. C. Chung, and 
R. A. M. S. Nazareth, 
Phys. Rev. Lett. $\bf49$, 536 (1982).
\bibitem{Goncalves1997} M. Goncalves, S. de Pina, D. A. Lima {\it et al.}, 
Phys. Lett. B $\bf406$, 1 (1997).
\bibitem{Depp2013b} A. Deppman, {\it et al.}, Phys. Rev. C $\bf88$, 
024608 (2013).
\bibitem{Depp2013c} A. Deppman, E. Andrade-II, V. Guimar\~aes,
G. S. Karapetyan, and N. A. Demekhina, Phys. Rev. C $\bf87$, 
054604 (2013).

\bibitem{Depp2013a} A. Deppman, {\it et al.}, Phys. Rev. C $\bf88$, 
064609 (2013).
\bibitem{Strecker} M. Strecker, R. Wien, P. Plischke, and W. Scobel, 
Phys. Rev. C  {\bf 41}, 2172 (1990).

\end{thebibliography}
\end{document}